\documentclass[%
 reprint,
superscriptaddress,
 amsmath,amssymb,
prb,
]{revtex4-2}

\usepackage{amsmath}
\usepackage{graphicx}
\usepackage{dcolumn}
\usepackage{hyperref}
\usepackage{textcomp}
\usepackage[dvipsnames]{xcolor}
\usepackage{bm}

\begin{document}

\preprint{APS/123-QED}

\title{Programmable electrical coupling between stochastic magnetic tunnel junctions }

\author{Sidra Gibeault}
\author{Temitayo N. Adeyeye}
\author{Liam A. Pocher}%
\author{Daniel P. Lathrop}%
\affiliation{%
Institute for Research in Electronics and Applied Physics, University of Maryland, College Park, MD, USA
}%


\author{Matthew W. Daniels}
\author{Mark D. Stiles}
\author{Jabez J. McClelland}
\author{William A. Borders}
\author{Jason T. Ryan}
\affiliation{
Physical Measurement Laboratory, National Institute of Standards and Technology, Gaithersburg, MD, USA
}%
\author{Philippe Talatchian}
\author{Ursula Ebels}
\affiliation{%
Univ.~Grenoble Alpes, CEA, CNRS, Grenoble INP, SPINTEC, 38000 Grenoble, France
}%

\author{Advait Madhavan}
\affiliation{%
Institute for Research in Electronics and Applied Physics, University of Maryland, College Park, MD, USA
}%
\affiliation{
Physical Measurement Laboratory, National Institute of Standards and Technology, Gaithersburg, MD, USA
}%


\begin{abstract}

Superparamagnetic tunnel junctions (SMTJs) are promising sources of randomness for compact and energy efficient implementations of probabilistic computing techniques. Augmenting an SMTJ with electronic circuits, to convert the random telegraph fluctuations of its resistance state to stochastic digital signals, gives a basic building block known as a probabilistic bit or $p$-bit. Though scalable probabilistic computing methods connecting $p$-bits have been proposed, practical implementations are limited by either minimal tunability or energy inefficient microprocessors-in-the-loop. In this work, we experimentally demonstrate the functionality of a scalable analog unit cell, namely a pair of $p$-bits with programmable electrical coupling. This tunable coupling is implemented with operational amplifier circuits that have a time constant of approximately 1~{\textmu}s, which is faster than the mean dwell times of the SMTJs over most of the operating range. Programmability enables flexibility, allowing both positive and negative couplings, as well as coupling devices with widely varying device properties. These tunable coupling circuits can achieve the whole range of correlations from $-1$ to $1$, for both devices with similar timescales, and devices whose time scales vary by an order of magnitude. This range of correlation allows such circuits to be used for scalable implementations of  simulated annealing with probabilistic computing.

\end{abstract}

\maketitle




\section{\label{sec:intro}Introduction}



Probabilistic computing is becoming increasingly popular given its applicability to many classes of optimization problems and its amenability to hardware acceleration~\cite{lucas2014ising}. The constraints of a given problem are encoded as interactions between Ising spins, which vary randomly to explore configuration space. Conventional implementations either generate random numbers in a centralized location and distribute them to the computational cores~\cite{grimaldi2023evaluating}, or use microprocessors with digital-to-analog converters in the feedback loop to mediate interactions~\cite{borders2019integer,si2023energy}, neither of which are conducive to scaling. In this work, we use low barrier magnetic tunnel junctions (MTJs) to implement the Ising spins and analog circuits to mediate direct interactions between them, leading to scalable analog implementations of probabilistic computing kernels. 

The central objects in probabilistic computers are probabilistic bits, also known as $p$-bits~\cite{camsari2017stochastic}. Unlike a conventional Boolean bit, which takes only deterministic \textbf{1} and \textbf{0} values, a $p$-bit uses its room-temperature instability to encode a probability value. This is particularly useful in applications such as energy-based machine learning models~\cite{kaiser2022hardware,aadit2021computing}, combinatorial optimization~\cite{aadit2022massively}, and quantum simulation~\cite{lucas2014ising,chowdhury2023accelerated}. These applications map their problem statements into graphs of interconnected binary nodes with variable interaction strengths~\cite{aadit2022massively}. The parameters of the problem at hand are encoded in the interaction strengths, while the collective configuration of the binary nodes encodes the current proposed solution. The interaction strengths define an effective energy for each global configuration~\cite{pervaiz2018weighted}. If the state of each node is probabilistic, the configuration evolves with time and can explore configuration space according to a Boltzmann distribution, tending to spend more time in configurations with lower effective energy. Optimal solutions or solutions within the required error margin can be found using simulated annealing~\cite{bertsimas1993simulated}, in which the interaction strengths are collectively increased in a way analogous to lowering the temperature so that the configuration spends more time in the lowest energy state. 

Implementing such calculations on conventional computing systems can be time-consuming and inefficient. For large problems, the interconnection weights are necessarily stored on off-chip memory, while iterative calculations of the solution are performed on chip~\cite{jouppi2017datacenter}. Since conventional computing systems have separate locations for memory and computation, most of the performance limitations arise from having to move data between memory and computation centers~\cite{wulf1995hitting} (the so-called von Neumann bottleneck). If stochastically fluctuating $p$-bits are used as the nodes, a large binary state space can be searched efficiently, in terms of both area and energy cost. Large arrays of low-barrier versions of MTJs would be particularly appealing implementations of $p$-bits provided they can be coupled by locally engineered, electronically tunable interactions between them. This would allow memory (interaction strengths) and computation (accumulation and randomness generation) to be co-located, promising significant performance boosts~\cite{sebastian2020memory}. 

Magnetic tunnel junctions are nanoscale magnetic devices that consist of two ferromagnetic layers separated by a thin insulator~\cite{jinnai2020scaling}. The insulating layer is thin enough for electrons to tunnel from one ferromagnetic layer to another. The resistance seen by the tunneling electrons depends on the relative orientation of the magnetizations of the ferromagnetic layers~\cite{julliere1975tunneling}. There are two stable configurations of the magnetizations, namely parallel and antiparallel (denoted as P and AP). The device resistance, also known as the tunnel magnetoresistance (TMR), is smaller if the magnetizations are parallel to each other than if they are antiparallel~\cite{moodera1995large}. The relative orientations between layers can be changed by applying an external magnetic field across the device or passing a current through the device~\cite{slonczewski1996current,ralph2008spin}.

MTJs have various properties that make them useful as memory devices when integrated into modern complementary metal oxide semiconductor (CMOS) chips~\cite{worledge2022spin,garello2019manufacturable}. Their stable configurations are used to encode binary values: for example, the parallel state may encode a value of \textbf{0} and the antiparallel state a value of \textbf{1}. When used in storage applications~\cite{lee20191gbit,aggarwal2019demonstration} requiring long retention times ($\approx 10$ years), the energy barrier between stable configurations is engineered to be $\geq 40~kT$, where $k$ is Boltzmann's constant and $T\approx 300$~K is room temperature. While MTJ-based magnetic random access memory (MRAM) is a promising candidate for various memory and unconventional computing applications~\cite{grollierNeuromorphicSpintronics2020,aggarwal2019demonstration,jung2022crossbar,dong20181,golonzka2018mram}, reducing the energy barrier reduces exponentially the retention time of the MTJ~\cite{kanai2021theory,rippard2011thermal}.

For energy barriers less than $14~kT$, thermal fluctuations at room temperature switch the device between its stable configurations~\cite{bapna2017current} at time scales faster than a millisecond. In this regime, devices are said to be superparamagnetic tunnel junctions (SMTJs). A current applied across the device allows its fluctuating resistance state to be read out as a random telegraph voltage signal and simultaneously controls the relative time that the device spends in each of its stable states~\cite{rippard2011thermal}. The ability to generate tunable random signals makes SMTJs useful for applications where cheap, energy-efficient randomness is required~\cite{daniels2020energy,shukla2023,schnitzspan2023nanosecond,vodenicarevic2017low}. Conventional CMOS sources of randomness such as linear feedback shift registers are only pseudo-random and require comparatively large area and energy budgets~\cite{daniels2020energy}.

Though proposals of large-scale probabilistic computing systems using SMTJs as $p$-bits have been reported in literature \cite{aadit2022massively,chowdhury2023accelerated,borders2019integer}, practical implementations are still limited to a handful of devices. The largest demonstrations use up to 80 $p$-bits and perform tasks such as integer factorization~\cite{borders2019integer}, invertible logic gates~\cite{camsari2017stochastic}, learning Boolean functions~\cite{kaiser2022hardware} and travelling salesman problems~\cite{si2023energy}. Larger scale demonstrations are limited by practical factors such as a lack of integrated platforms for exploration, yield and repeatability failures in the fabrication process, the need for an external magnetic field, and differences in time scales between memory and computation. External magnetic fields in particular, which are required to set devices in the superparamagnetic regime in many experiments, are inefficient and impractical in commercial applications. 

Mismatch of timescales is an especially important issue. In probabilistic computing, the states of the nodes are sampled and that set of sampled states determines the input controlling the state of each of the nodes. A key requirement for correct operation of probabilistic circuits is that determining the input for each node must complete before the the next sample is taken, so that the subsequent sample is representative of the input~\cite{camsari2017stochastic}. For free running interactions, this requirement means that the time scales of the interaction circuits (which can become relatively complex), must be much smaller than the timescales of nodes themselves. Previous demonstrations~\cite{borders2019integer,kaiser2022hardware,si2023energy} measure the device state and perform the calculations of the accumulated interactions with a microprocessor in the loop. This approach works because the devices have switching times in the tens of milliseconds or more, which is enough for thousands of microprocessor cycles to perform computations digitally. 

Although using a microprocessor in the loop is useful to demonstrate the functionality and expressibility of probabilistic computing, it is impractical for scalable implementations. Recent papers report simple ways of directly coupling devices. When two SMTJs are connected in parallel~\cite{talatchian2021mutual,phan2022electrical} or in series~\cite{mizrahi2016synchronization,schnitzspan2023electrical} to a voltage source via a tunable resistor, the current flowing through one device depends on the state of the other. Though such approaches can scale theoretically~\cite{daniels2023neural}, practical problems such as line resistance, device-to-device variability, and noise quickly overwhelm linear circuit solutions as the number of devices grows. Moreover, such simple connections between devices are not enough to implement the large scale interactions required to build practical systems. 

The fastest and most energy-efficient implementations would consist of CMOS integrated with SMTJs in the back end of the line~\cite{jung2022crossbar}. Such implementations would involve analog stages mediating interactions between SMTJ-based $p$-bit unit cells. In this work, we take a step in that direction with commercial off-the-shelf analog circuits. We experimentally demonstrate the functionality of a unit cell coupling circuit between $p$-bits, with direct analog interaction between field-free SMTJs that have operational time scales in the tens to hundreds of microseconds. The speeds of the analog computation circuits themselves are much faster, on the order of 1~{\textmu}s, allowing the devices to faithfully couple to each other. The degree of coupling can be tuned by varying the interaction strengths, while the uncoupled probabilities of the devices can be independently set by adjusting bias currents. These two features, coupled with a programmable gain circuit, provide the necessary ingredients to perform optimization based on simulated annealing. This unit cell can be readily adapted to couple more than two devices, making it a useful stepping stone towards integration of arrays of devices. 

The rest of this paper is organized as follows. Section~\ref{sec:circuit} describes the coupling circuit and discusses how this circuit can be used to perform simulated annealing. Section~\ref{sec:engineering} contains a description of the experimental setup and the details of the specific SMTJs used. Section~\ref{sec:Results} presents the experimental results of coupling devices of various time scales. Sec.~\ref{sec:MarkovModelling} describes a Markov model that accurately describes the switching time scales of the two-SMTJ coupled system, and Sec.~\ref{sec:Discussion} discusses scalability of the coupling circuit.

\section{\label{sec:circuit}SMTJ Interactions}

\begin{figure*}
    \centering
    \includegraphics[width=17.5cm]{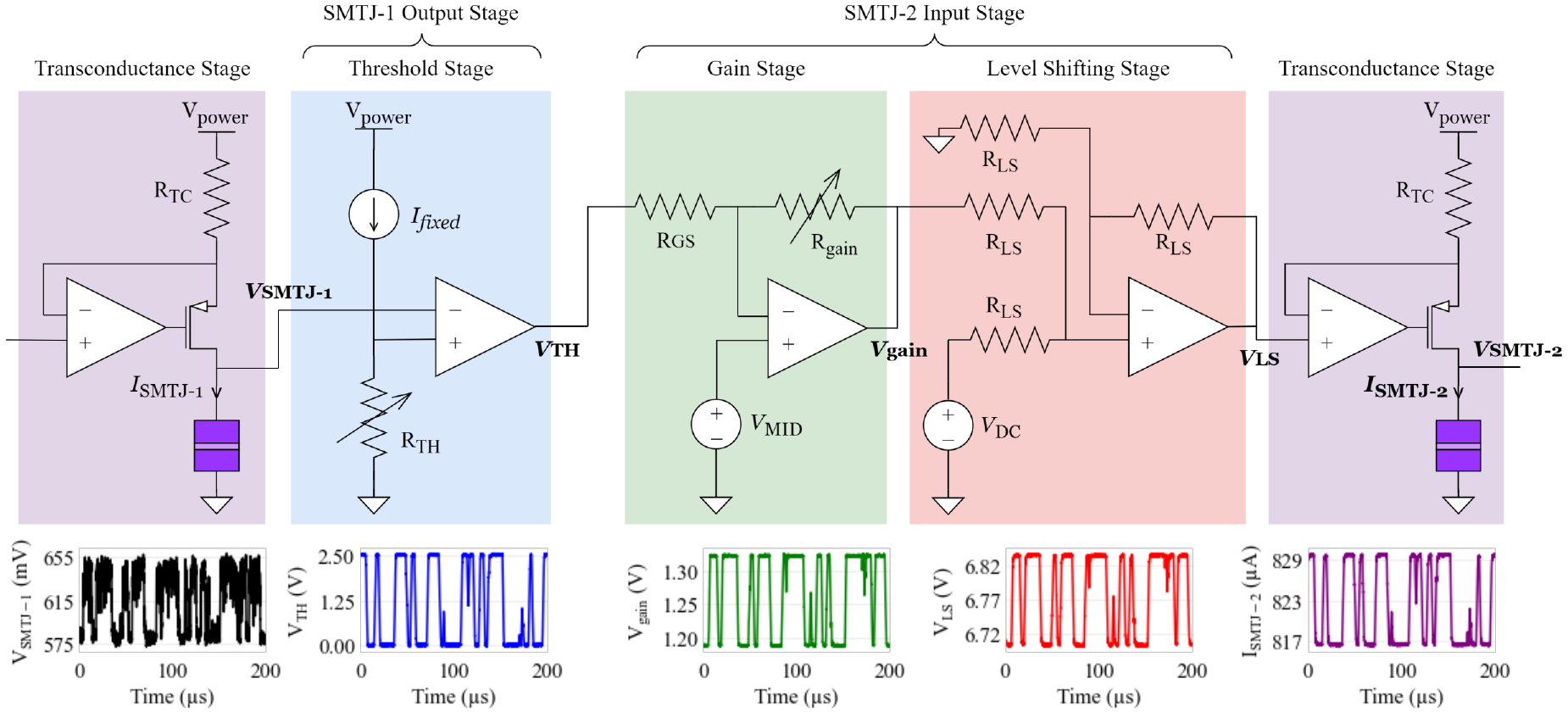}
    \caption{Uni-directional SMTJ interaction circuit. Two copies of this circuit are used, one in each direction, to achieve bi-directional coupling. The SMTJ voltage gets sensed by a threshold stage, which converts the stochastic fluctuations to a digital output signal using a programmable reference voltage set by $I_{fixed}$ and $R_{TH}$. This digital signal is passed through a gain stage, which adjusts its magnitude while retaining the mid voltage level. The gain $G = R_{gain}/R_{GS}$ is programmable. The level shifting stage adjusts the mid voltage level while retaining the voltage difference between high and low levels. $V_{LS}$ is adjustable and controls the size of the shift. The shifted voltage is converted by a transconductance stage to a fluctuating current, which facilitates influencing the next SMTJ. In this example, the current in the next SMTJ is an inverted version of the SMTJ state, but both positive and negative coupling are possible with our circuit. In this example, $R_{TH} = 615~\Omega$, $I_{fixed} = 1~mA$, $G = 0.05$, $V_{MID} = 1.25~V$, and $V_{DC} = 5.5~V$.}
    \label{fig:ckt-diagram}
\end{figure*}

\subsection{\label{circuit_a}Interaction circuit }
Our coupling circuit allows two SMTJs to mutually influence the current through each other. The circuit senses the resistive state of one SMTJ and adds a differential current to a second SMTJ depending to the state of the first. The strength of the coupling can be tuned by adjusting the magnitude of the differential current. We make the coupling bi-directional, meaning that both SMTJs are influencing the current through each other, by using two independent, uni-directional interaction circuits. The uni-directional interaction circuit, shown in Fig.~\ref{fig:ckt-diagram}, consists of several operational amplifier stages that enable the raw, noisy SMTJ voltage to influence a second SMTJ. These signal conditioning stages, which can be classified as either input or output stages, have specific functions in the interaction pipeline.  

The gain stage, which is the first stage in the input pipeline, receives its input from the last stage of the output pipeline of the previous device as shown in Fig~\ref{fig:ckt-diagram}. Its function is to set the strength of the coupling by adjusting the signal swing. This is done by setting the value of the $R_{GS}$ resistor, which together with the $R_{gain}$ resistor determines a programmable gain value $G = R_{gain}/R_{GS}$. The gain stage typically reduces the voltage spread of its digital input signal to a desired level in order to correctly influence the device it controls. In section~\ref{sec:Discussion}, we describe how the gain stage can be augmented to support multiple SMTJ inputs for scaled implementations.  The next stage in the input pipeline is the level shifting stage which shifts the mean voltage of the gain-stage signal to some pre-determined adjustable level. This level is typically appropriate for biasing the influenced device to the equal probability state. Therefore, when no input signal is received by the gain stage, a static current biases the influenced device to its maximum entropy state.  The transconductance stage, which is the last stage of the input pipeline, turns the shifted voltage signal into a proportional two-level current, which is then input to the next SMTJ. Since the differential part of the current is proportional to $G$, the strength of the coupling from the previous SMTJ is adjusted through the gain. 

The output pipeline consists of a single stage which is the threshold stage. The noisy signal from the SMTJ is converted into a digital output signal based on a programmable reference voltage which is controlled by $I_{fixed}$ and $R_{TH}$. There is also a polarity selection in the threshold stage that can be used to invert the state of the device, effectively producing positive or negative coupling. The gain, SMTJ bias and threshold reference voltage are programmed via digital potentiometers in this implementation.

\subsection{\label{circuit_b}SMTJ model and simulated annealing}
The modified N\'eel-Brown model \cite{PhysRev.130.1677,rippard2011thermal} describes how the current flowing through an SMTJ affects its probability of being in either magnetoresistive state. The mean dwell time in each state at a current $I$ is
\begin{equation}\label{eq:neel-brown-model}
    \tau_{\pm}(I) = \tau_{0} \exp\left[-\frac{\Delta E}{kT}\left(1 \mp \frac{I - I_{0}}{I_c}\right)\right]
\end{equation}
where $\tau_{0}$ is the characteristic time scale, $\Delta E$ is the energy barrier, $T$ is the temperature, here nominally room temperature, $k$ is the Boltzmann's constant, $I_c$ is the critical current, and $I_{0}$ is the current at which the device spends equal time in each state. Here $\tau_{+}$ is the mean dwell time in the antiparallel (AP) state and $\tau_{-}$ is the mean dwell time in the parallel (P) state. 

The probability of finding the device in either state can be written as the fraction of total time spent in that state
\begin{eqnarray}
    P_\pm(I) &=& \frac{\tau_{\pm}(I)}{\tau_{\pm}(I) + \tau_{\mp}(I)} \nonumber \\
    &=& \frac{1}{1+\exp\left[\mp\frac{2\Delta E}{kT}\left(\frac{I-I_{0}}{I_c}\right)\right]}.
\end{eqnarray}
By changing the current applied to an SMTJ, the probability of being in either resistive state also changes.

Our implementation of two balanced unidirectional interaction ciruits can be thought of as the simplest unit cell of an Ising machine~\cite{mohseni2022ising} and is capable of performing simulated annealing. In simulated annealing, the objective is to find the ground state of spins in an interconnected graph that minimizes the global energy as described by the Ising Hamiltonian~\cite{lucas2014ising}. For our simple circuit, the Hamiltonian is determined by the weights of the two-edge graph (uni-directional interactions) connecting the SMTJs. The joint state of the SMTJ-pair that minimizes the energy is the desired solution. In this way, simulated annealing can be performed by starting with a gain $G$ of  0 and gradually increasing it. 

During simulated annealing, the probability of being in the $i$-th joint state, e.g. (P,P), (AP,AP), (P,AP), (AP,P), is determined by a Boltzmann distribution
\begin{align}\label{eq:effective_energy}
    P_i(T) =& \frac{1}{Z(T)}{\exp\left[-\frac{E_i(G)}{kT}\right]},
\end{align}
where $E_i(G)$ is the effective energy of the $i$-th joint state at gain value $G$ and $Z(T) = \sum_i \exp\left[{-E_i(G)}/{kT}\right]$. This effective energy can be linearly related to a model energy $E_i(G) = G\Tilde{E}_i$. Here, $\Tilde{E}_i=J_{12}S_1^i S_2^i$ where $S_n^i$ is the state of SMTJ $n$ in configuration $i$ and $J_{12}$ is the nominal coupling strength for that pair.
 
Rewriting the probability in terms of the model energy we get the following relation:
\begin{align}
    P_i(T_\text{eff}) =& \frac{1}{Z}{\exp\left[\frac{-\tilde{E}_i}{kT_\text{eff}}\right]}. \label{eq:model_temp}
\end{align}
Here, $T_\text{eff}=T/G$ is the effective temperature of the model. 

The process of simulated annealing proceeds as follows. Initially, the two-device circuit, which is at room temperature, has gain equal to zero. Equation \ref{eq:model_temp} shows that even though the system is at room temperature, the model temperature diverges in the sense that all joint SMTJ states are equally probable; the devices are effectively uncoupled. As the gain increases, the interaction circuit causes the devices to become correlated. The model temperature drops and the system spends more time in lower-energy joint states as prescribed by Boltzmann statistics. In conventional simulated annealing, the system would finally converge to a solution with the lowest energy; in our case, at maximal gain, the system still has enough energy from room temperature to explore low-energy degenerate states. SMTJs have a barrier breakdown current \cite{chavent2020multifunctional, ma2021microwave} that imposes a limitation on how high the gain can be set without destroying the SMTJs. Higher breakdown currents would allow us to lower the temperature further for better convergence, but infinite gain (zero temperature) would still not be possible.

\section{\label{sec:engineering}Experimental setup}

\begin{figure}
    \centering
    \includegraphics[width=8.5cm]{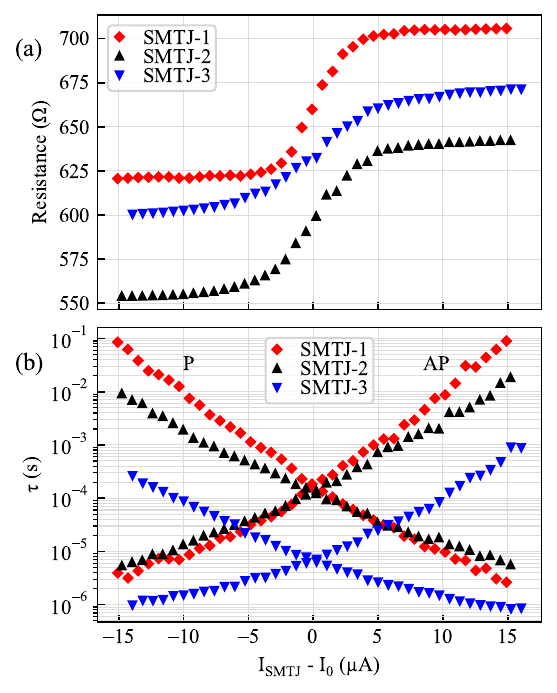}
    \caption{Characteristics of the SMTJs used in our experiments. (a) Average resistance as a function of applied current. (b) Mean dwell time in the P (negative slope curves) and AP (positive slope) states as a function of applied current. SMTJ-1 and SMTJ-2 have similar timing characteristics, while SMTJ-3 switches an order of magnitude faster on average. The current at which the SMTJ is in the AP state 50~\% of the time is denoted as $I_{0}$, which is 0.95~mA, 0.90~mA, and 1.00~mA for SMTJ-1, SMTJ-2, and SMTJ-3, respectively. Statistical uncertainties are smaller than the plotting symbols.}
    \label{fig:smtj-characteristics}
\end{figure}

To demonstrate the coupling of two SMTJs, we use perpendicular magnetic tunnel junctions with the following stack sequence: Si base / SiO2 / TaN / [Co(0.5)/Pt(0.2)]6 / Ru(0.8)/ [Co(0.6)/Pt(0.2)]3 / Ta(0.2) / Co(0.9) / W(0.25) / CoFeB(1) / MgO(0.8) / CoFeB(1.4) / W(0.3) / CoFeB(0.5) / MgO(0.75) / Ta(150) / Ru(8). The numbers in parentheses refer to layer thicknesses in nanometers, while the numbers beside square brackets show the bilayer repetitions. These devices, almost circular in shape, have approximate diameters of 150~nm. They exhibit a TMR near 120~\% at room temperature and possess a resistance-area product of $10~\Omega${\textmu}m$^2$. These devices were initially developed as multifunctional magnetic tunnel junctions, useful for both non-volatile memory and radio frequency (RF) applications \cite{chavent2020multifunctional, ma2021microwave}. Barrier breakdown has been observed in these devices at currents around 1~mA (corresponding to voltages around $0.7$~V).

Field-free stochastic magnetization fluctuations have been observed in these perpendicular MTJ devices for a large range of diameters; for details see Ref.~\cite{el2022radiofrequency}. These devices allow for field-free fluctuations for two reasons. First, at the CoFeB thickness of 1.4~nm, the perpendicular magnetic anisotropy \cite{dieny2017perpendicular} nearly cancels the self-demagnetization field, leaving only a small energy barrier. Second, the fringing field from the fixed layer is small enough that the energies of the parallel and antiparallel configurations can be aligned with relatively small external magnetic fields or moderate currents.
Field-free operation of the coupling experiments ensures that operation remains practical and commercially viable. However, in the absence of a field, the current required to set these SMTJs to 50~\% probability of being in each state is quite high ($\approx~0.95$~mA), necessitating operation close to barrier failure and suggesting refinement of the device design for future applications.

Figure~\ref{fig:smtj-characteristics}(a) shows the average resistances of the SMTJs used in our experiments as a function of applied current. $I_0$ here is defined as the current at which the probabilities of being in either device state are equal. The resistances are low and almost independent of current for currents much below $I_0$ because the SMTJs are stable in the parallel state. As the current through the device approaches $I_0$, the increasing spin-transfer torque causes the device to start switching stochastically, spending increased time in the antiparallel state. This corresponds to the sharp increase in resistance between $-3$~{\textmu}A to 3~{\textmu}A. As currents are further increased, the rate of resistance increase slows down since the device spends all of its time in the antiparallel state.  

Figure~\ref{fig:smtj-characteristics}(b) shows the mean dwell times of the devices in the parallel and antiparallel states as a function of applied current. The parallel state dwell times are monotonically decreasing, while the antiparallel state dwell times are monotonically increasing. This is again by virtue of lower currents stabilizing the parallel state, while large currents stabilize the antiparallel state. Although the resistance values of SMTJ-1 and SMTJ-2 shown in  Fig.~\ref{fig:smtj-characteristics}(a) are considerably different, the difference in resistance does not affect the coupling dynamics, because the coupling circuit allows us to set the threshold voltage and $I_0$ separately for each SMTJ. We perform experiments with this pair of devices as the ``ideal'' case, treating them as identical due to their similar time scales over the current range of Fig.~\ref{fig:smtj-characteristics}(b). On the other hand, SMTJ-3, whose resistance is between SMTJ-1 and SMTJ-2, is roughly an order of magnitude faster than the other two devices. Experiments done with SMTJ-2 and SMTJ-3 study the effect of coupling devices that have different time scales. 

The printed circuit board (PCB) designed for this experiment performs multiple functions. It receives inputs from the probe station on which the SMTJ chip resides via coaxial cables connected to ground-signal (GS) probes. Each stage of the interaction circuit has outputs that can be connected to an oscilloscope for debugging and data collection. Peripheral bias and control inputs required to adjust the $I_{0}$ currents, threshold voltages, and interaction strengths are controlled by a microcontroller via a standard communication protocol. Both the microcontroller and the oscilloscope are connected to a host computer that performs all of the control operations, data collection, and analysis in software.

There is approximately 1~{\textmu}s of propagation delay in our circuit; when one SMTJ switches, it takes $\approx 1$~{\textmu}s for the current through the other SMTJ to change in response. This delay makes our circuit sub-optimal for high-speed applications; however, more specialized amplifiers could be chosen to reduce the propagation delay. As discussed in Sec.~\ref{sec:Discussion}, the delay would also be significantly reduced in an integrated circuit compared to a PCB. 

We conduct the experiment as follows. First, both SMTJs are provided their respective $I_0$ currents to ensure that the probabilities of being in each state are equal. Then the gain is increased, increasing the effect that each SMTJ has on the other. Voltage-time traces are collected from each SMTJ at specific increments of gain. We then conduct statistical analyses on these voltage-time traces to demonstrate that increased gain concomitantly increases the degree of influence between the SMTJs. 

\section{\label{sec:Results}Results}

\begin{figure}
    \centering
    \includegraphics[width=8cm]{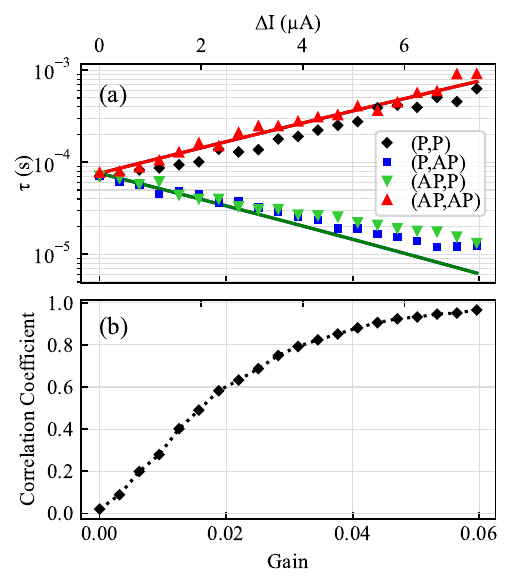}
    \caption{Demonstration of positive coupling between SMTJ-1 and SMTJ-2, which have similar switching time scales. (a) Log-linear plot of mean dwell times in each of the four joint states the pair of SMTJs can be in. Positive coupling causes the SMTJs to spend more time in the same state (both in P or both in AP) and less time in opposite states with increasing gain. Solid lines are calculated mean joint dwell times using the Markov model described in Section~\ref{sec:MarkovModelling}. (b) Pearson correlation coefficient between SMTJ voltage-time traces as gain is increased from 0 to 0.06. Both top and bottom axes apply to both panels. Statistical uncertainties are smaller than the plotting symbols.}
    \label{fig:noninv-identical-corr}
\end{figure}

\begin{figure}
    \centering
    \includegraphics[width=8cm]{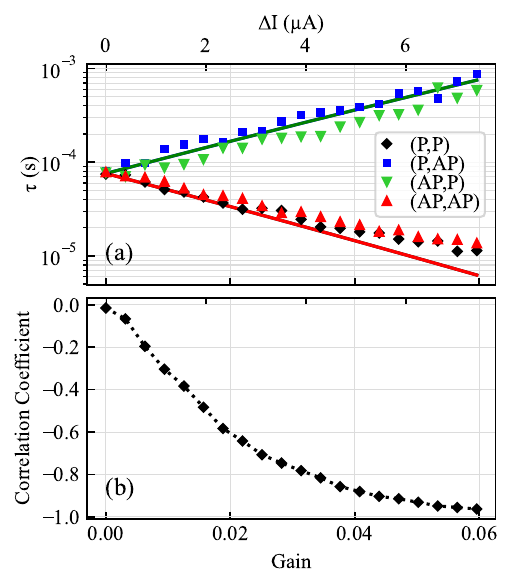}
    \caption{Demonstration of negative coupling between SMTJ-1 and SMTJ-2, which have similar switching time scales. (a) Log-linear plot of mean dwell times in each of the four joint states the pair of SMTJs can be in. Negative coupling causes the SMTJs to spend more time in opposite states (one in P and the other in AP) and less time in the same state with increasing gain. Solid lines are calculated mean joint dwell times using the Markov model described in Section~\ref{sec:MarkovModelling}. (b) Pearson correlation coefficient between SMTJ voltage-time traces as gain is increased from 0 to 0.06. Both top and bottom axes apply to both panels. Statistical uncertainties are smaller than the plotting symbols.}
    \label{fig:inv-identical-corr}
\end{figure}

The SMTJs become more strongly coupled as the stochastic time between one SMTJ switching and the second SMTJ switching in response is reduced. If we consider the four joint states of the two-SMTJ system (P,P), (P,AP), (AP,P), and (AP,AP), we can define joint dwell times to be the mean dwell times in each of these four joint states. With positive coupling, when the system is in the (P,P) or (AP,AP) states, the currents through both SMTJs are such that there is a very low probability of either SMTJ switching out of its current state. We thus expect the dwell times in joint states (P,P) and (AP,AP) to increase with gain, and the dwell times in joint states (P,AP) and (AP,P) to decrease with gain, which is the behavior seen in Fig.~\ref{fig:noninv-identical-corr}(a). In the negatively coupled case, the joint states (P,AP) and (AP,P) are stabilized, so the dwell times in these states increase with gain as seen in Fig.~\ref{fig:inv-identical-corr}(a). 

Note that in Fig.~\ref{fig:noninv-identical-corr}(a), the (P,AP) and (AP,P) dwell times decrease linearly until around $G = 0.03$, where the curve begins to flatten out. This is a sampling artifact. As the gain increases, the shorter dwell times start to approach the sampling times. As these times become closer some short-time events get missed, making the mean dwell times of the shorter events artificially larger, but also combining two of the long-time events, making those mean dwell times artificially longer as well. Computing dwell times requires saving entire time traces, making faster sampling prohibitive. However, these issues do not affect the calculation of the equal-time correlation coefficients between the two tunnel junctions, which is the main quantity of interest as a measure of the joint interaction between the two SMTJs.

The Pearson correlation coefficient can be used to quantify the similarity between the (digitized) SMTJ voltage-time traces for both positive and negative coupling (Fig.~\ref{fig:noninv-identical-corr}(b) and Fig.~\ref{fig:inv-identical-corr}(b), respectively). A Pearson correlation coefficient of $1$ indicates high similarity between the two voltage-time traces, namely that the SMTJs spend a lot of time in the correlated joint states (P,P) and (AP,AP), and very little time in the anti-correlated joint states (P,AP) and (AP,P). A Pearson correlation coefficient of $-1$ indicates that the SMTJs are spending a lot of time in the joint states (P,AP) and (AP,P), and very little time in the joint states (P,P) and (AP,AP). A Pearson correlation coefficient of $0$ indicates equal time in correlated and anti-correlated states. In the positive coupling experiment, the Pearson correlation starts at $0$ and approaches $1$ in the high gain limit. The negative coupling experiment has opposite dominant joint states to the positive coupling case, which results in the Pearson correlation approaching $-1$.

Previous methods of coupling between SMTJs \cite{talatchian2021mutual,phan2022electrical,schnitzspan2023electrical} which facilitate interaction via linear circuit elements have yielded a maximum Pearson correlation coefficient of $0.35$. Such a low maximal correlation coefficient, coupled with the fact that these methods use a single resistor to control the interaction between their constituent SMTJs, make such approaches difficult to scale for simulated annealing tasks. On the other hand, our analog interaction circuit has a pair of independently programmable resistors for each pairwise set of SMTJs. It also features a global gain resistor that selects the effective temperature of the problem. Together, these features result in a Pearson correlation coefficient that approaches $1$. This implies that our method is more viable for simulated annealing, where it is important to implement programmable pairwise interactions and be able to lower the effective temperature as much as possible.

\begin{figure}
    \centering
    \includegraphics[width=8cm]{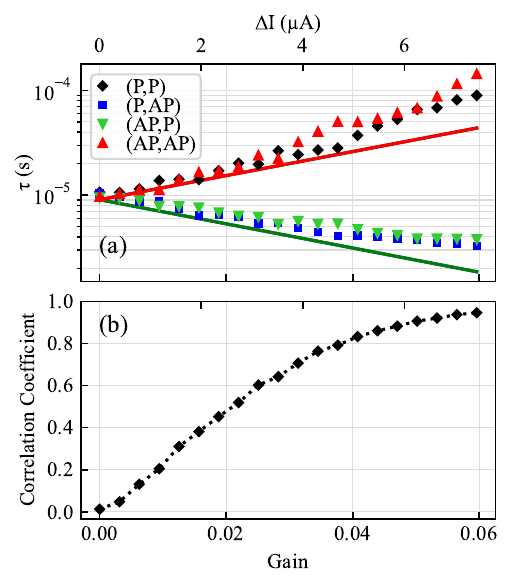}
    \caption{Demonstration of positive coupling between SMTJ-2 and SMTJ-3, which have different switching time scales. (a) Log-linear plot of mean dwell times in each of the four joint states the pair of SMTJs can be in. Positive coupling causes the SMTJs to spend more time in the same state (both in P or both in AP) and less time in opposite states with increasing gain. Solid lines are calculated mean joint dwell times using the Markov model described in Section~\ref{sec:MarkovModelling}. (b) Pearson correlation coefficient between SMTJ voltage-time traces as gain is increased from 0 to 0.06. Both top and bottom axes apply to both panels. Statistical uncertainties are smaller than the plotting symbols.}
    \label{fig:noninv-diff-corr}
\end{figure}

The results discussed so far have demonstrated an analog circuit that facilitates tunable coupling between SMTJs with similar time scales. Figure~\ref{fig:noninv-diff-corr} shows that the circuit successfully couples devices with very different time scales. When the positive coupling experiment is repeated with SMTJs of different time scales, the Pearson correlation and joint dwell times, plotted in Fig.~\ref{fig:noninv-diff-corr}, are very similar to the results in Fig.~\ref{fig:noninv-identical-corr}, aside from the joint state time scales being overall faster. In the next section, we examine how this difference in switching time scales affects the system's relaxation to equilibrium.

\section{\label{sec:MarkovModelling}Markov Modelling}

Since the coupling between SMTJs in our experiments is facilitated entirely by analog circuitry, device properties such as TMR and $I_0$ have less of an impact on the coupling strength or on the behavior of the coupled system than in recent SMTJ coupling work~\cite{talatchian2021mutual,schnitzspan2023electrical}. However, the difference in time scales of the coupled SMTJs does have a sizeable impact on the dynamics of the coupled system. To examine this effect, we can look at the eigenvalues of the Markov transition matrix for this system. This is a $4\times 4$ matrix containing the transition rates between each of the four joint states  (P,P), (P,AP), (AP,P),
and (AP,AP), which we index as states 00, 01, 10, and 11, respectively. 

\begin{figure}
    \centering
    \includegraphics[width=8.5cm]{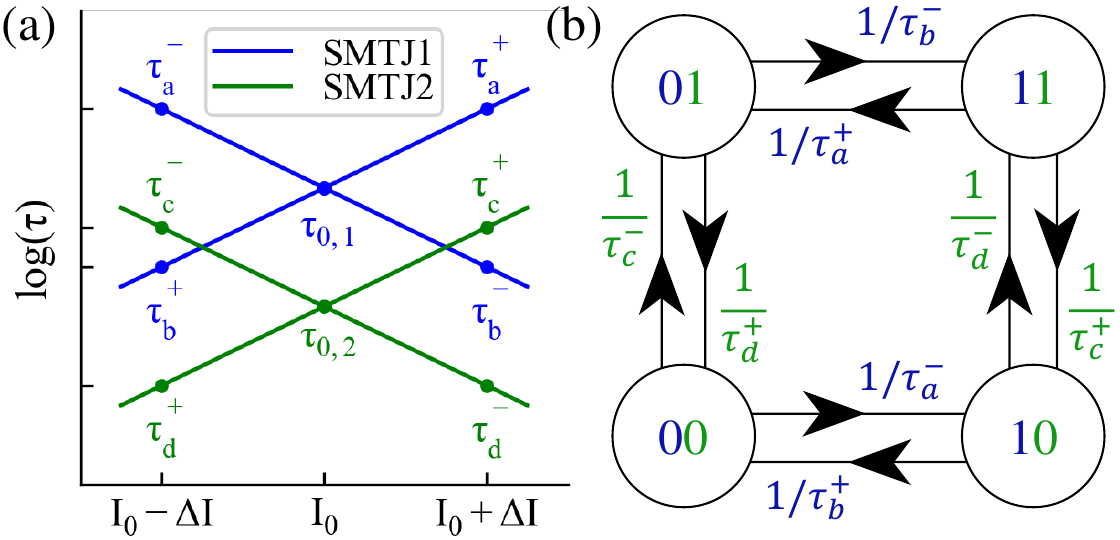}
    \caption{(a) Conceptual mean dwell time plot (similar to Fig.~\ref{fig:smtj-characteristics}(b)) showing the inverse switching rates in the two-SMTJ coupled system. Positive slope lines represent dwell times in the AP state, while negative slope lines represent dwell times in the P state. Note the log scale on the vertical axis. The mean dwell times in the P and AP states are equal at $I_0$, and these dwell times are labeled as $\tau_{0,1}$ and $\tau_{0,2}$. The remaining labels are dwell times at either $I_0 + \Delta I$ or $I_0 - \Delta I$, which are used to write the transition rates from one state into another. (b) Markov transition diagram for the two-SMTJ coupled system. The switching rate out of each state is effectively a single-device switching rate, which is dependent on the state of the other SMTJ due to coupling.}
    \label{fig:markov-tau-plot}
\end{figure}

We consider transitions where both SMTJs switch simultaneously (for instance, a switch from (P,P) to (AP,AP)) to have probability zero; all nonzero transition rates in our system represent single-SMTJ switching events. The switching rate of one SMTJ is dependent on the state of the other SMTJ due to the coupling; in the positive coupling case, when SMTJ-1 is in the P state, the current through SMTJ-2 will be $I_0 - \Delta I$. When SMTJ-1 is in the AP state, the current through SMTJ-2 will be $I_0 + \Delta I$. Referring to Fig.~\ref{fig:markov-tau-plot}(a), we can then determine which inverse rate is the correct one for each of the eight possible transitions from one state into another. For example, $1/\tau^+_a$ is the rate for the first SMTJ to switch from the antiparallel state to the parallel state when the second SMTJ is in the antiparallel state, and $1/\tau^-_a$ is the rate for the first SMTJ to switch from the parallel state to the antiparallel state when the second SMTJ is in the parallel state. 

The transition rate diagram is shown in Fig.~\ref{fig:markov-tau-plot}(b), with Fig.~\ref{fig:markov-tau-plot}(a) showing the relative locations of the inverse switching rates with respect to each other on a log-linear plot of mean dwell time against current. The notation in Fig.~\ref{fig:markov-tau-plot} can be simplified with the assumption that both devices have the same slopes for the current dependences, in which case $\tau_{n}^+ = \tau_{n}^- = \tau_n$, where $n = \{a, b, c, d\}$. Since we are interested in examining how the difference in time scales of the two devices impacts the coupled dynamics, we represent the difference in time scales by a scaling factor $r$ on $\tau_{0}$ via
\begin{equation}
    \tau_{0,2} = \frac{1}{r}\tau_{0,1}.
\end{equation}
Note that $r$ will always be greater than 1 since we will always choose SMTJ1 to be the slower device. Rewriting the $\tau_n$s in terms of $\tau_0,1$ yields
\begin{equation}\label{eq:tau_simplifications}
    \begin{split}
        \tau_{a} &= g\tau_{0,1}\\
        \tau_{b} &= \frac{1}{g}\tau_{0,1}
    \end{split}
    \quad\quad\quad
    \begin{split}
        \tau_{c} &= \frac{g}{r}\tau_{0,1}\\
        \tau_{d} &= \frac{1}{gr}\tau_{0,1},
    \end{split}
\end{equation}
where $g = \exp\left[B\Delta I\right]$ represents the change in the dwell times as a function of gain ($G$), since $\Delta I$ is linearly related to the gain. We would like to emphasize that although the gain $G$ is proportional to the current increment $\Delta I$, $g$ is an exponential function of both $G$ or equivalently $\Delta I$. For the sake of clarity, we have chosen to make $g$ for the two SMTJs equal. This implies that $B$ for the two SMTJs are equal, which would make the slopes of the lines in Fig.~\ref{fig:markov-tau-plot}(a) equal. 

To generate a Markov model for our coupled system, we refer to Appendix B of Ref.~\cite{talatchian2021mutual}. The Markov transition rate matrix $\mathbf{M}$ can be directly generated from Fig.~\ref{fig:markov-tau-plot}(b) using the simplifications in Eq.~\ref{eq:tau_simplifications}. The steady state probability distribution is given by the eigenvector corresponding to the zero eigenvalue of $\mathbf{M}$, namely
\begin{equation}
    v_0 = \left(1, \frac{1}{g^2}, \frac{1}{g^2}, 1 \right).
\end{equation}
Note that this eigenvector, which encodes the (unnormalized) steady state distribution, is only a function of $g$, and the probability of being in the two disfavored states decreases quadratically with the value of $g$. Note that $g$ itself depends exponentially on the gain.

In addition to this zero eigenvalue, there are three negative eigenvalues. The one closest to zero, provides the slowest rate of decay into the steady state distribution. In order to see how selecting SMTJs with different time scales affects the coupled dynamics, we can look at the eigenvalue that produces the term which decays most slowly,
\begin{gather}
    \lambda_1 = \frac{1}{\tau_{0,1}}  \frac{-b+\sqrt{-16g^2r+b^2}}{2g} \\
    b = (g^2+1)(r+1), \nonumber
\end{gather}
which depends on both $g$ and $r$. So although the final distribution is governed by $g$ alone, the time scale of the coupled system's relaxation to equilibrium increases not only with gain but also with a higher ratio of SMTJ time scales. 


We use this model to predict the mean joint dwell times during the coupled experiments. These model predictions are plotted as solid lines on the experimental dwell time plots in Section~\ref{sec:Results} (Fig.~\ref{fig:noninv-identical-corr}(b), Fig.~\ref{fig:inv-identical-corr}(b), and Fig.~\ref{fig:noninv-diff-corr}(b)). To apply the model to our experiment, we drop the assumption that $B$ is the same for both SMTJs, reflecting the different slopes in Fig.~\ref{fig:smtj-characteristics}. The result of this is that each SMTJ has its own $g$ value, and the steady state distribution becomes dependent on $r$ in addition to $g$. We fit lines to the characteristic dwell time plots of Fig.~\ref{fig:smtj-characteristics} to obtain $B$, which we use to calculate $g$ for each SMTJ. $\tau_{0,1}$ and $r$ were also determined from Fig.~\ref{fig:smtj-characteristics} for each pair of SMTJs. The negative reciprocals of the diagonal elements of $\mathbf{M}$ give the dwell times in each of the four joint states for a given $\Delta I$. 



\section{\label{sec:Discussion}Discussion}

\begin{figure}
    \centering
    \includegraphics[width=7cm]{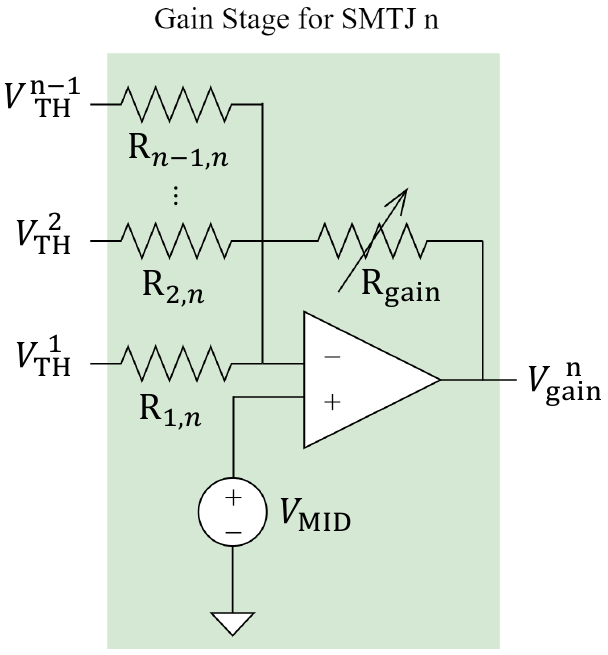}
    \caption{Interaction circuit gain stage in a network of n SMTJs. Each SMTJ's gain stage receives voltage inputs from the threshold stages of the other $n-1$ SMTJs. This way, the output of the gain stage contains the total influence from all other SMTJs in the network. This modification is a simple method of scaling the proposed coupling method up to large values of $n$. The number of operational amplifiers scales as $n$ while the number of resistors scales as $n^2$.}
    \label{fig:gain-stage-scaling}
\end{figure}

The analog circuits used to facilitate coupling in this paper can be scaled up for larger bench-top experiments or integrated implementations with simple modifications. The threshold stage, used to convert noisy SMTJ signals to clean digital levels can be implemented by skewed inverter circuits with a digitally programmable threshold, or an analog latch-based sense amplifier design. Such circuits have operating speeds $<150$~ps and are composed of $<10$ transistors~\cite{wicht2004yield,capra2018digitally}. Transconductance amplifiers for level shifting bias voltages and adding differential currents can also be practically implemented at nanosecond time scales. SMTJ switching speeds can be significantly increased before CMOS input and output signal conditioning circuits become latency bottlenecks. 

In a network with many SMTJ circuits such as the one described here, the gain stage of each interaction circuit can be modified to accept inputs from each of the other SMTJs in the network. The gain stage will then be a summing amplifier circuit, with rows of resistors feeding into the amplifier, as shown in Fig.~\ref{fig:gain-stage-scaling}. This scalability opens the possibility of creating an Ising machine to solve realistically large problems.  The number of operational amplifiers in the interaction circuits grows linearly with the number of nodes in such an Ising machine, while the number of resistors needed for the modified gain stage grows quadratically if the nodes are all-to-all connected. These quadratic circuits can be implemented with crossbars of resistive switches. The push for building hardware platforms for matrix multiplication has led to the development of low-latency, energy-efficient large-scale crossbars. CMOS-integrated resistive RAM crossbars in submicron nodes ($22$~nm and $130$~nm) have recently been demonstrated on such applications with arrays as large as $1024\times512$. When scaled to representative kernel sizes ($256\times256$), such crossbars incur latencies on the order of $\approx1$~{\textmu}s to $10$~{\textmu}s~\cite{wan2022compute,hung2021four}. The crossbar arrays implementing the all-to-all connections are the the most area, energy, and latency-intensive aspects of larger-scale designs. Solving larger problems will require codesign of devices, architectures and algorithms. Ising Hamiltonians with large connectivity matrices will be partitioned into crossbar kernels that solve a subset of the global problem. Additional nodes to reduce the connectivity requirements for particular problems~\cite{aadit2022massively} would be needed, augmented with specific circuits to handle communication between kernels.

Since $p$-bit based Ising machines operate on the principle that the programmable interaction circuits should operate at speeds faster than the nodes themselves, microsecond switching SMTJs may be sufficient for scaled integrated implementations. However, the SMTJs used in this study are far from optimal. Estimates of enhanced SMTJ properties for integrated designs involve lower operating currents~\cite{hu20212x}, faster operating speeds~\cite{schnitzspan2023nanosecond}, and larger TMR~\cite{hu20212x}. Larger TMR enables easier CMOS readout, while lower operating currents are essential for energy efficiency. The high energy cost of large static currents can be amortized by fast switching circuits that produce more random bits per unit time and hence current, but becomes difficult to justify when faster operation is not beneficial for directly coupled designs. 


\section{\label{sec:Acknowledgements}Acknowledgements}

This work was funded by the National Institute of Standards and Technology, National Science Foundation and Agence nationale de la recherche. SG, TA, LP, DPL and AM acknowledge support under NSF grant number CCF-CISE-ANR-FET-2121957. AM also acknowledges support under the NIST Cooperative Research Agreement Award No. 70NANB14H209 through the University of Maryland. PT and UE acknowledge support under the ANR StochNet Project award No.~ANR-21-CE94-0002-01. The authors would like to thank Steve Moxim for help with experimental setup and Frank Mizrahi for valuable feedback. The authors acknowledge J.~Langer and J.~Wrona from Singulus Technologies for the MTJ stack deposition and N.~Lamard, R.~Sousa, L.~Prejbeanu as well as the Upstream Technological platform PTA, Grenoble, France for the device nanofabrication. 

\nocite{*}

\bibliography{references}

\end{document}